\documentclass[a4paper]{article}

\usepackage[english]{babel}
\usepackage[utf8]{inputenc}
\usepackage[T1]{fontenc}
\usepackage{authblk}
\usepackage{csquotes}

\usepackage[
    style=vancouver,
    giveninits,
    sorting=none,
    uniquename=init,
    citestyle=numeric-comp,
    ]{biblatex}
\AtEveryBibitem{%
   \clearfield{urldate}
   \clearfield{language}
   \ifentrytype{online}{}{
    \clearfield{url}
    
  }
}
\DeclareFieldFormat*{journaltitle}{\mkbibitalic{#1}}

\usepackage[a4paper,top=3cm,bottom=2cm,left=3cm,right=3cm,marginparwidth=1.75cm]{geometry}

\usepackage{amsmath, amssymb, amsthm, amsfonts}
\usepackage{graphicx}
\usepackage{wrapfig}
\usepackage[font=small,labelfont=bf]{caption}
\usepackage{subcaption}
\usepackage{placeins} 
\usepackage{url}
\usepackage{siunitx}
\usepackage[dvipsnames]{xcolor}
\usepackage{soul}
\sethlcolor{yellow}
\usepackage[colorlinks=true, allcolors=blue]{hyperref}

\newcommand{\Rey}{\mathrm{Re}}
\newcommand{\Cd}{C_d}
\newcommand{\Cdbar}{\overline{C}_d}
\renewcommand{\vec}[1]{\mathbf{#1}}

\title{Fluid--structure interactions of bristled wings: \\The trade-off between weight and drag}
\author{Yuexia Luna Lin, Matteo Pezzulla\footnote{Current affiliation: Slender Structures Group, Department of Mechanical and Production Engineering, {\AA}rhus University, {\AA}rhus, Denmark}, Pedro M. Reis\footnote{To whom correspondence should be addressed: pedro.reis@epfl.ch}}

\affil{\footnotesize{Ecole Polytechnique F\'{e}d\'{e}rale de Lausanne (EPFL)\\
Flexible Structures Laboratory\\
Lausanne, CH-1015, Switzerland
}}
\begin{document}
\maketitle

\begin{abstract}
The smallest flying insects often have bristled wings resembling feathers or combs.
We combined experiments and three-dimensional numerical simulations to investigate the trade-off between wing weight and drag generation.
In experiments of bristled strips, a reduced physical model of the bristled wing, we found that the elasto-viscous number indicates when reconfiguration occurs in the bristles.
Analysis of existing biological data suggested that bristled wings of miniature insects lie below the reconfiguration threshold, thus avoiding drag reduction.
Numerical simulations of bristled strips showed that there exist optimal numbers of bristles that maximize the weighted drag when the additional volume due to the bristles is taken into account.
We found a scaling relationship between the rescaled optimal numbers and the dimensionless bristle length.
This result agrees qualitatively with and provides an upper bound for the bristled wing morphological data analyzed in this study.
\end{abstract}

\section*{Keywords}
Insect flight; bristled wings; fluid-structure interactions, reconfiguration, drag of porous structures 

\section{Introduction}

Insects are some of the finest flyers in nature, and insect flight has been the subject of abundant scientific research~\cite{pringle1957, dudley2002} and bio-inspired engineering~\cite{ellington1999, liu2016, de_croon2022}.
Most studies on insect flight focus on macroscopic insects with membraned wings, such as flies, moths, and dragonflies~\cite{sane2003, wang2005, dickinson2016, shyy2016, salami2019, nakata2020}. 
However, there is a myriad of miniature flying insects with body size less than \SI{1}{\milli\meter}~\cite{sane2016, polilov2016}; some as small as a few hundred microns, e.g., a species of fairyflies called \textit{Tinkerbella nana}~\cite{huber2013}.
In contrast to macroscopic insects, the morphology, behavior, and flight of miniature insects and the fluid-structure interactions (FSI) of their wings are much less well-understood, despite having long fascinated scientists~\cite{thompson1961, horridge1956, cheer1987}.

\begin{figure}[t]
        \centering 
        \includegraphics[width=\textwidth, keepaspectratio]{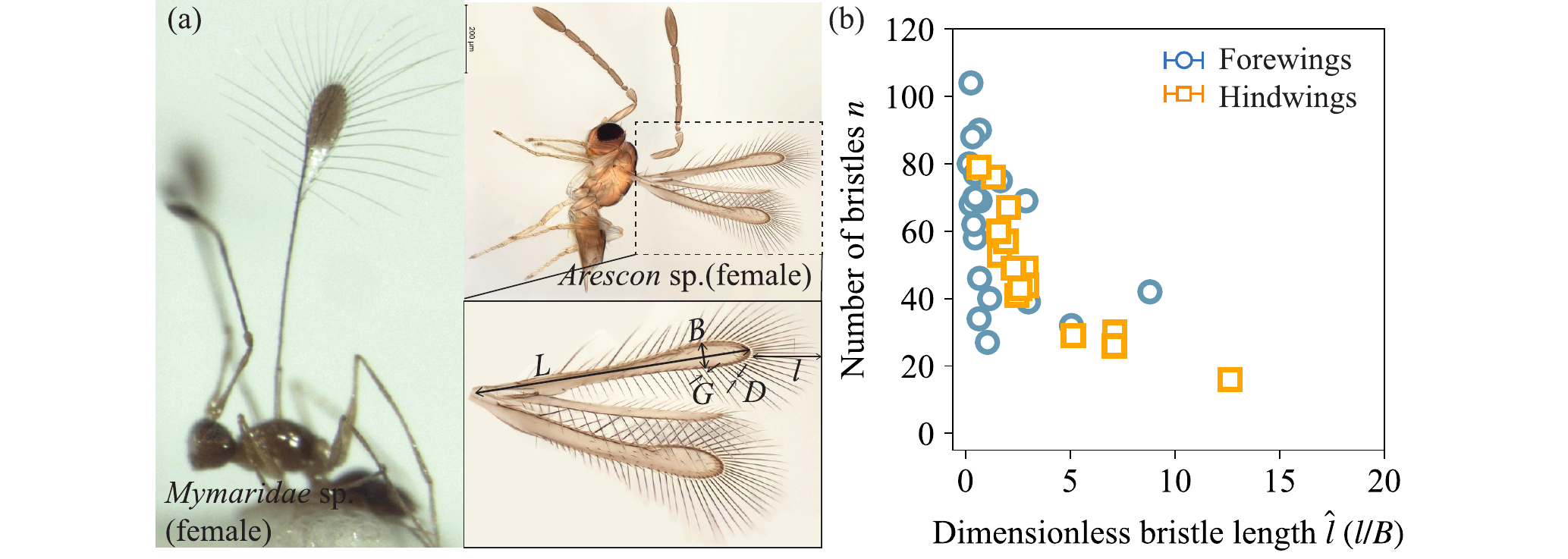}
        \caption{\textbf{Bristled wings of miniature insects.}
        (a) Photographs of miniature insects with bristled wings, adapted from Wikimedia \cite{mymaridaeWiki, arescon2011}.
        Bottom right: a close-up image of the wings of \textit{Arescon} sp. specimen \cite{arescon2011}. The geometric parameters are indicated: wing length $L$, width $B$, (average) bristle length $l$, bristle diameter $D$, and inter-bristle gap $G$. The number of bristles is denoted as $n$. 
        (b) Number of bristles $n$ plotted against the ratio $l/B$ in insects, where $l$ is the average bristle length. Measurements gathered or estimated from refs.~\cite{jones2016, lin2007, huber2006, huber2007, huber2008, huber2013, cheng2018, farisenkov2022, jiang2022} are tabulated in Tables S1 and S2 in the supplementary information.}
        \label{fig:definition}
    \end{figure}

Instead of membrane-covered wings, miniature flying insects are often donned with feather-like wings~\cite{mymaridaeWiki, arescon2011, lapina2021}, with  filaments emanating from a central wing pad (Fig.~\ref{fig:definition}a).
These wings are collectively called bristled wings.
Studies of a wide range of miniature insects~\cite{huber2006,lin2007, huber2007, huber2008,huber2013, jiang2022, farisenkov2022} provide detailed data on the number, shape, and micro-structure of the bristles, as well as the dimensions of the central wing pads.
In addition, the flight characteristics of several species of bristled-winged insects have been quantified by high-speed videos,
allowing for analysis and numerical simulations of flight kinematics and aerodynamics at extremely small size scales~\cite{cheng2018, zhao2019, lyu2019, yavorskaya2019, kolomenskiy_aerodynamic_2020, farisenkov2022}.
These studies revealed that instead of the horizontal sweeping stroke used commonly by larger insects, miniature insects engage bristled wings in a figure-of-8 stroke cycle~\cite{jones2015, lyu2019, farisenkov2020, engels_flight_2021, farisenkov2022, shen2022}.
The Reynolds number (Re) of the bristled wing is usually estimated based on the wing chord length ($\Rey_c$) and has been reported to be $\Rey_c {=} \mathcal{O}(1)-\mathcal{O}(10)$~\cite{horridge1956, weis-fogh1975, engels_flight_2021}.

To investigate the aerodynamic benefits of bristled wings in miniature insects, many studies have combined theoretical, experimental, and computational methods to make progress.
Santhanakrishnan et al.~\cite{santhanakrishnan_clap_2014}, and Jones et al.~\cite{jones2016} used two-dimensional (2D) approximations of bristles in low Re ($\Rey_c$<10) simulations to study clap-and-fling motions found in smaller insects~\cite{weis-fogh1975, ellington1999, dudley2002, miller2009}.
Leakiness is a common parameter to characterize bristled structures, defined as the ratio between the actual flux moving between a pair of bristles and the flux if the fluid were inviscid~\cite{cheer1987}. The studies mentioned before \cite{santhanakrishnan_clap_2014, jones2016} have found that as the leakiness increases, the force required to separate the wings after the clap reduces.  

Drag and lift force production by bristled wings in linear and rotational translations and clap-and-fling movements have also been characterized~\cite{sunada2002, lee_plate_2017, kasoju_leaky_2018, ford_aerodynamic_2019, kolomenskiy_aerodynamic_2020, kasoju_interspecific_2021}. These studies confirmed that, at low Re, bristled wings enjoy aerodynamic advantages compared to their membraned counterparts because they experience only slightly decreased lift force while greatly reducing wing weight.
In particular, the study by Kasoju et al.~\cite{kasoju_leaky_2018} on comb-like wing models with clap-and-fling actuation ($\Rey_c${=}5-15) confirmed previous numerical studies \cite{santhanakrishnan_clap_2014, jones2016} and further asserted that, compared to membraned wings, bristled ones had enhanced drag-to-lift ratio.
Using a similar experimental setup, Ford et al.~\cite{ford_aerodynamic_2019} and Kasoju et al.~\cite{kasoju_interspecific_2021} showed that aerodynamic forces are only weak functions of the  geometric parameters of the wing, such as the number of bristles and inter-bristle gap.
The authors concluded that optimizing aerodynamic force generation during clap-and-fling might not drive the interspecific variations in bristled wing morphology.

Flow near the bristles and around the wing is difficult to visualize experimentally~\cite{kasoju_leaky_2018, ford_aerodynamic_2019}.
As such, computational models have been widely used to gather data on the dynamics of the shear layers between bristles due to flapping \cite{lee_unsteady_2018} and due to perturbation to steady steaming flow \cite{lee_gusty_2021}, and the changing force production and aerodynamic efficiency due to variations in bristle parameters~\cite{lee_optimal_2020, kolomenskiy_aerodynamic_2020, wu_aerodynamics_2021, shen2022}.

The power consumption of wing flapping, intimately related to the inertial load of the wings, is a crucial factor for flight efficiency.
Numerical studies have shown that bristled wings have a large advantage over membraned ones in reducing wing weight, thus the power consumption of flapping, especially in lower Re regimes~\cite{farisenkov2020, engels_flight_2021}. 
The more lightweight designs of bristled wings tend to feature fewer and thinner bristles,  leading necessarily to a competition between the reduction in aerodynamic force generation and the reduction of wing weight.
However, this trade-off has rarely been examined in previous studies exploring optimal configurations of bristled wings or natural selection pressures that drive the evolution of their morphology~\cite{lee_optimal_2020, wu_aerodynamics_2021, jiang2022}.
Furthermore, these numerical studies have not been compared quantitatively with biological data.

Here, we combine experiments and three-dimensional (3D) FSI simulations to investigate the common morphological trends of bristled wings (Fig.~\ref{fig:definition}).
Via a dimensional analysis, the experimental and biological data are compared to address the question of whether the bristles on insect wings reconfigure during flight. Our results  suggest avoiding reconfiguration is a common feature of a wide variety of bristled wings.
Leveraging validated simulations, we investigate the trade-off between the increased drag generation and increased wing weight due to additional bristles.
We propose a drag coefficient, weighted by the additional bristle volume, to assess bristled wing performance.
We uncover a power-law scaling  between rescaled optimal bristle numbers determined from the simulations and a dimensionless bristle length.
Furthermore,  the simulation results provide an upper bound below which lies the biological data.

Our paper is organized as follows. In Section~\ref{sec:prob_def}, we define the questions we aim to address and introduce the reduced model we employed in our study. 
Next, in Section~\ref{sec:mats_methods}, we present the sample-fabrication procedure, experimental apparatus and protocol, and numerical simulation methods.
In Section~\ref{sec:results}, we report our results from experiments, computations, and direct comparisons with existing biological data.
Finally, in Section~\ref{sec:dis}, we discuss some limitations of this work  and conclude with a brief summary and outlook for future work in Section~\ref{sec:conclusion}.

\section{Problem definition}
\label{sec:prob_def}

We seek to characterize the trade-off between the increased drag generated by a bristled wing and the increased weight due to additional bristles. We will make use of experiments and simulations to quantify the wing's drag coefficient and its dependence on the geometric parameters.
We start by defining  the relevant geometric parameters of a biological bristled wing as shown in Fig.~\ref{fig:definition}a.
Denote the length of the central wing pad $L$, its width $B$,  the number of bristles $n$, the length of the bristles $l_i, i{=}1,2,...,n$, its diameter $D$, and the inter-bristle distance $G$, measured from edge to edge between adjacent bristles. 
We also define the average bristle length $l$ and a dimensionless bristle length $\hat l {=} l/B$.
The values of $D$, $G$ were taken directly from Ref.~\cite{jones2016}, while other parameters are measured using ImageJ~\cite{schneider2012} in images of bristled wings from published studies~\cite{lin2007, huber2006, huber2007, huber2008, huber2013, farisenkov2022}.
In each image, to estimate $l$, we manually measure 5 bristles, 2 from the region on the sides of the wing pad and 3 from the region at the tip of the wing pad.
The maximum and minimum bristle lengths were also obtained from the set of 5 values.
The biological data we surveyed is documented in Tables S1 and S2 in the supplementary information (SI) and presented in Fig.~\ref{fig:definition}b; the number of bristles $n$ decreases as the dimensionless bristle length $\hat l$ increases. In Section~\ref{sec:result:sim_data}, we will revisit this dataset with insight from numerics.

We consider a reduced model of the wing, referred to as bristled strip, to create an experimental system that simplifies the biological problem.
As shown in Fig.~\ref{fig:setup_exp_and_sim}a, the bristled strip has a rectangular central core, and its bristles have uniform length, diameter, and inter-bristle spacing.
Mimicking the parametrization of the biological wing, we denote the length of the central core $L$, its width $B$, and its thickness $t$.
The parameters $l, D, G, n$ have the same geometrical meaning as in a biological wing.
The fabrication of our bristled strips is detailed in Section~\ref{sec:mats_methods:sample}.
We design experiments (see Section~\ref{sec:mats_methods:exp}) to quantify the drag force $F_\mathrm{drag}$ generated by the bristled strips using a linear translation motion, with constant velocity $U$ and an angle of attack of $90^\circ$ (Fig.~\ref{fig:setup_exp_and_sim}b).

The smallest flying insects have been reported to flap their wings in a figure-of-8 cycle, where the wings are engaged at a large angle of attack for much of the stroke cycle~\cite{farisenkov2020, engels_flight_2021, farisenkov2022, shen2022}.
The aerodynamic force responsible for generating lift during the flapping motion arises primarily from the drag force acting in opposition to the wing motion.
Often referred to as a drag-based mechanism, the figure-of-8 flapping gait contrasts with the lift-based mechanism observed in larger insects, which involves horizontal  back-and-forth flapping to generate aerodynamic forces perpendicular to wing motion~\cite{jones2015, lyu2019}. The essential role of drag in the low Re flight justifies our selection of a model system centered on the drag experienced by the bristled strip translated at a $90^\circ$ angle of attack.

The Re based on the central core length is $\Rey_L{=} \rho U L / \mu$, where $\rho$ and $\mu$ are the mass density and dynamic viscosity of the fluid, and $U$ is the constant plunging velocity.
The Reynolds numbers $\Rey_L$ and $\Rey_c$ in our experiments are on the same order of magnitude; we choose $\Rey_L$ because it more conveniently reflects the experimental conditions.
The Stokes drag $F_\mathrm{drag}$ of an object with a characteristic length $L$ and fluid velocity $U$ scales as $\mu U L$~\cite{batchelor1967}, and the Stokes drag coefficient can be defined as
\begin{equation}
    \Cd = \frac{F_\mathrm{drag}} { \mu U L}.
    \label{eq:cd}
\end{equation}

Several previous works on bristled wings~\cite{cheer1987, barta_creeping_2006, lee_optimal_2020, jones2015, jones2016, wu_aerodynamics_2021} have focused on the aerodynamics of actual biological wings or that of simulated  2D reduced models, e.g., quantifying drag and lift coefficients of wings in rotational movements and at various angles of attack.
In this study, we will focus on the drag generation of a reduced representation of the bristled wing.
Experimentally, we aim to quantify how the Stokes drag coefficient of a bristle strip depends on the number, length, and diameter of the bristles.
Through validated simulations, we aim to explore the role of geometric parameters on the weight and drag of the bristles strips and gain insight into morphological trends observed in biological bristled wings.


\section{Materials and Methods}
\label{sec:mats_methods}

In this section, we will describe the experimental and computational methodology followed throughout the study. We will present the fabrication of bristled strip samples in Section~\ref{sec:mats_methods:sample}, the experimental apparatus and protocols in Section~\ref{sec:mats_methods:exp}, and numerical simulations in Section~\ref{sec:mats_methods:num}.

\subsection{Sample fabrication}
\label{sec:mats_methods:sample}

\begin{figure}[ht]
        \centering 
        \includegraphics[width=\textwidth, keepaspectratio]{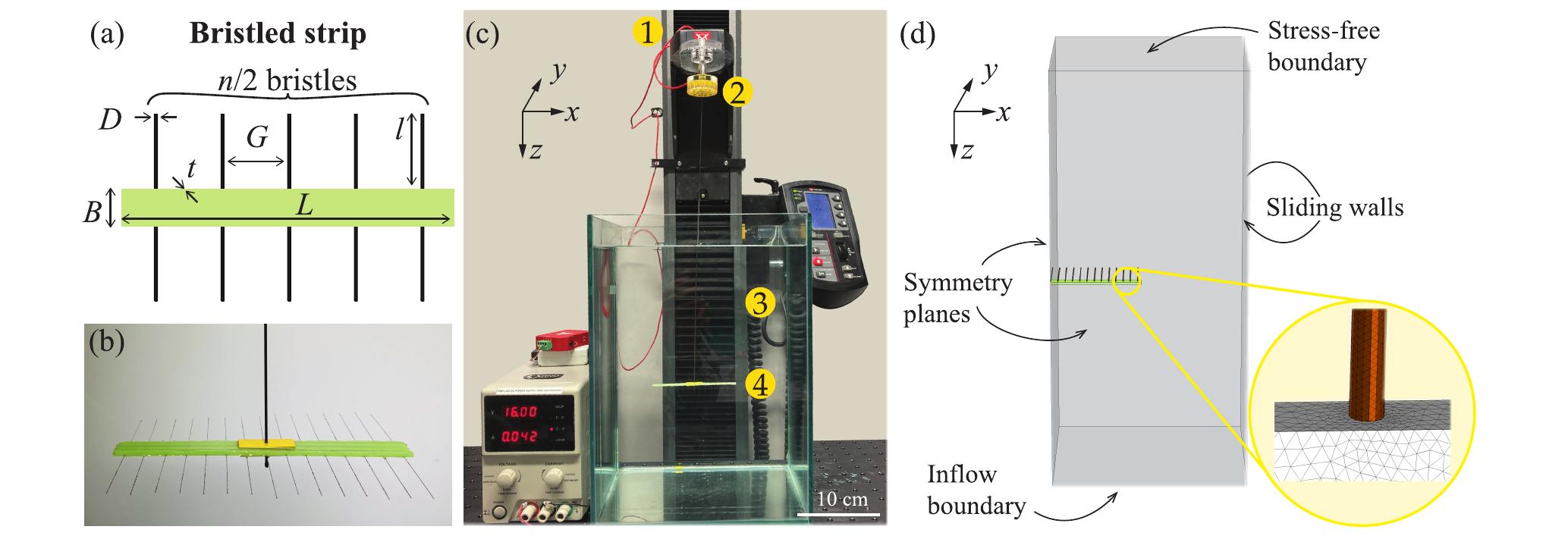}
        \caption{\textbf{Experimental and simulation configurations.} (a) Schematic of a bristled strip. (b) Photograph of a bristled-strip specimen suspended by the sample holder.
        (c) A Photograph of the experimental apparatus, including (1) a universal testing machine, (2) a \SI{50}{\gram} Futek load cell attached to the sample holder, (3) a bath of 1000 cSt silicone oil, and (4) a bristled-strip sample.
        (d) Schematic of a representative COMSOL simulation. Only a quarter of the system is simulated due to symmetry, resulting in two symmetry planes. Inflow and stress-free boundary conditions are prescribed at the inlet (bottom) and outlet (top).
        Constant velocity $-\vec{U}{=}-U\hat{z}$ is prescribed at the physical walls, and no-slip boundary condition is applied to all other solid surfaces.
        The zoomed-in image shows the meshing details  near the bristled strip.}
        \label{fig:setup_exp_and_sim}
    \end{figure}

A schematic and a photograph of a bristled strip sample are shown in Fig.~\ref{fig:setup_exp_and_sim}a and Fig.~\ref{fig:setup_exp_and_sim}b, respectively.
The central core of the bristled strips was made with a vinyl polysiloxane polymer (VPS) and reinforced with 2-3 Nitinol wires (Alfa Aesar, diameter \SI{0.58}{\milli\meter}) lengthwise to increase the bending stiffness so that the core does not deform during the experiments.
The acrylic mold used to cast the central core had a rectangular cavity of dimensions $L{=}\SI{10}{\centi\meter}, B{=}\SI{1}{\centi\meter}, t{=}\SI{0.1}{\centi\meter}$.
To place the Nitinol-wire bristles, we engraved evenly spaced grooves on the mold.
Given that VPS swells in silicone oil~\cite{pezzulla_2020, leroy-calatayud2022}, the different buoyancy forces due to changing volume affected the accuracy of force readings during the experiment.
Therefore, before using the strips in any experiment, we soaked them in a bath of silicone oil (BlueSil 47V1000, Siltech) for \SI{48}{\hour} to ensure saturation in swelling ~\cite{pezzulla_2020, leroy-calatayud2022}. Once swollen, the central core had dimensions of $\SI{10.5}{\centi\meter} {\times} \SI{1.05}{\centi\meter} {\times} \SI{0.15}{\centi\meter}$;
these values were used in simulations to reproduce the experimental results.

We fabricated three sets of samples, \textbf{A, B} and \textbf{C}, with identical central cores.
Nitinol wires with identical Young's modulus $E_n$ but varying diameters $D{=}0.58, 0.17, \SI{0.12}{\milli\meter}$ and lengths $l{=}3, 3, \SI{6}{\centi\meter}$ were used as the bristles for set \textbf{A, B, C}, respectively.
Within each set of samples, the number of bristles was varied: set \textbf{A} with $n{\in}[4,50]$,  set \textbf{B} with $n{\in}[6,30]$, and set \textbf{C} with $n{\in}[4, 40]$.
These parameters are summarized in Table~\ref{table:samples}.
Despite the care during  fabrication, there was some degree of misalignment of the bristles, i.e., bristles deviated slightly from the plane of the central core (visible as the slight blurring of the bristles in lateral view photographs in Fig.~\ref{fig:deformable_bristles}).
The deviation was especially evident in sample set \textbf{C}, with the longest and thinnest bristles.

\begin{table}[!ht]
    \centering
    \resizebox{0.5\textwidth}{!}{%
    \begin{tabular}{|c | c |c|c|}
    \hline
    Set ID & Diameter $D$ (mm) & Length $l$ (cm) & Number $n$  \\ \hline
    \textbf{A} & 0.58 & 3 & 4, 6, 8, 20, 30, 40, 50 \\ \hline
    \textbf{B} & 0.17 & 3 & 6, 10, 12, 16, 30 \\ \hline
    \textbf{C} & 0.12 & 6 & 4, 8, 20, 30, 40 \\ \hline
    \end{tabular}
    }
    \caption{Fabrication parameters for the bristled strip samples.}
    \label{table:samples}
\end{table}

\subsection{Experimental apparatus and protocol}
\label{sec:mats_methods:exp}

In Fig.~\ref{fig:setup_exp_and_sim}c, we show a photograph of the experimental apparatus.
A tank with square cross-section $\SI{23}{\centi\meter} {\times} \SI{23}{\centi\meter}$ contained 1000 {cSt} silicone oil (BlueSil 47V1000, Siltech); the fluid column height is $\SI{30}{\centi\meter}$.
A sample holder with an acrylic disc and a vertical Nitinol rod (Alfa Aesar, diameter \SI{1}{\milli\meter}) was connected to the load cell (Futek LSB200 50g), and then mounted on a universal testing machine (Instron 5943) and aligned concentrically with the tank.
 
Each strip sample mounted on the vertical rod was first set just below the surface ($\vec{z}{=}\vec{0}$) of the silicon oil bath, ensuring that no bubbles were entrained. Before initiating the plunging, the force signal,~$\vec{F}_\mathrm{drag} {=} - F_\mathrm{drag} \vec{\hat z}$, was recorded for at least $\SI{10}{\second}$ and the average value subtracted to tare the load cell.
The sample was then plunged into the bath at a constant velocity until $\vec{z} {=}\SI{25}{\centi\meter}\,  \vec{\hat z}$.
For each bristled strip, nine velocities were imposed, $\vec{U} {=} \{1, 5, 10, 15, 20, 25, 30, 35, 40 \}(\vec{\hat z})\unit[per-mode = symbol]{\milli\meter\per\second}$, each with five repetitions.
We denote the plunged distance $z{=}|\vec{z}|$ and the plunge velocity magnitude as $U{=}|\vec{U}|$.
The force signal $\vec{F}_\mathrm{drag}$ was recorded for the entire duration of the plunge path and for another 1-\SI{2}{\second} after the end of the plunge. Note that the load cell is one-directional and can only register forces along $\hat z$.
The strip is raised up to exactly the same initial position (near the free surface) between consecutive runs. The sample holder retains a coating of silicone oil, which is wiped manually with a tissue between runs.

The buoyancy and drag forces of the sample holder depend on $z$ and $U$. To isolate the drag force due to the bristle strip samples, the above procedure was applied to the sample holder alone~\cite{strong_2019}.
A time series of force data (buoyancy and drag combined) was obtained at each plunging velocity, averaged over five repetitions, and then subtracted from the measurements when a bristled strip was attached at the corresponding velocities.
This calibration procedure has been demonstrated to be effective for similar experiments in creeping flow~\cite{strong_2019}.
As we shall show in Section~\ref{sec:result:rigid_v_flex}, it is also valid for the Re regime in our experiments.

\subsection {Numerical simulations}
\label{sec:mats_methods:num}

Numerical simulations of bristled strips were conducted using COMSOL Multiphysics$^\text{\tiny{\textregistered}}$ 5.5 (henceforth COMSOL) in a configuration closely reproducing the experiments.
We only considered rigid bristled strips for reasons that will become clear in Section~\ref{sec:result:rigid_v_flex} and \ref{sec:result:scaling}.
A representative example of the simulation domain is shown in Fig.~\ref{fig:setup_exp_and_sim}d.
Due to the symmetry about the $x$-$z$ and the $y$-$z$ planes, only a quarter of the system was considered. 
Different from the experiments, the simulations neglect the sample holder (Nitinol rod), and the reference frame is centered at the bristled strip.
Therefore, instead of translating the sample at constant velocity $\vec{U}$, we prescribe a constant sliding velocity $-\vec{U}$ on the side walls of the fluid container.
At the upstream opening (bottom in Fig.~\ref{fig:setup_exp_and_sim}d), we used an inflow boundary condition (B.C.) with prescribed normal velocity and zero tangential velocity, and at the downstream opening (top in Fig.~\ref{fig:setup_exp_and_sim}d), a stress-free B.C.

The numerics were validated against experiments (sample sets \textbf{A} and \textbf{B} only).
The dimensions of the simulated domain $L_x {=} L_y {=} \SI{23}{\centi\meter}, L_z {=} \SI{25}{\centi\meter}$ matched those of the physical system.
Then, after the validation studies in Section~\ref{sec:result:sim_data}, we considered simulations in an unbound fluid domain to avoid wall effects.
To speed up the computation, we chose a sufficiently large fluid domain with $L_x^*{=}L_y^* {=} \SI{2.5}{\meter}$ and $L_z^*{=}\SI{1.4}{\meter}$.
We found $L_x^* (L_y^*{=}L_x^*)$ by first fixing $L_z$ and increasing $L_x$ and $L_y$ from \SI{0.25}{\meter} to \SI{5}{\meter}.
We defined the drag coefficient obtained in the largest domain $C_{d, \infty}$ and set $L_x^*$ to be the smallest value that yielded $99\%C_{d, \infty}$.
We set $L_z^*$ in a similar way.

We used tetrahedral elements and a three-level meshing strategy due to the challenge of separation of length scales: the fluid domain, the strip, and the bristles have drastically different characteristic sizes.
The bristled strip is at the finest level; the surface mesh size is bound by a minimum of ${D}/{12}$, where $D$ is the bristle diameter, and a maximum of $t$, the central core thickness.
At the intermediate level, we set a refinement region around the bristled strip with $L_x^r{=}L_y^r{=}L_z^r{=}\SI{0.3}{\meter}$ ($L_x^r{=}L_y^r{=}\SI{0.15}{\meter}, L_z^r{=}\SI{0.05}{\meter}$ for validation simulations). Here, the minimum element size is $D/12$, and the maximum is the width of the central core $B$.
Outside of the refinement region, we used the COMSOL preset option \texttt{Coarser} with a minimum and maximum element size of $\SI{0.05}{\meter}$ and $\SI{0.163}{\meter}$, respectively.
The option \texttt{Calibrate for fluid dynamics} was selected for all elements, dictating the default values of other element parameters (e.g., maximum element growth rate 1.25, curvature factor 0.8, and resolution of narrow regions 0.5). 
We used linear elements for both the velocity fields and the pressure, i.e., the \texttt{P1+P1} option for the discretization of the fluid, which is a good compromise between accuracy and computational speed.

For a typical simulation (e.g., a bristled strip from sample set \textbf{A} with $n{=}50$, cf. Table~\ref{table:samples}, in a domain with $L_x^*{=}L_y^* {=} \SI{2.5}{\meter}$ and $L_z^*{=}\SI{1.4}{\meter}$),
the complete mesh consisted of approximately 94000 domain elements, 36000 boundary elements, and 6700 edge elements.
Simulating this system using the \texttt{Laminar flow} module and solving the stationary Navier--Stokes equations with $U{=}1\,\unit[per-mode = symbol]{\milli\meter\per\second}$ $(\Rey_L{=}0.1)$ took approximately 6.5 GB of RAM and \SI{30}{\second} of wall time on a workstation with 48 cores (two Intel Xeon (R) Platinum 8268 CPUs).

\section{Results}
\label{sec:results}


\subsection{Reconfiguration-dependent fluid drag}
\label{sec:result:rigid_v_flex}

\begin{figure}[!htb]
        \centering 
        \includegraphics[width=\textwidth, keepaspectratio]{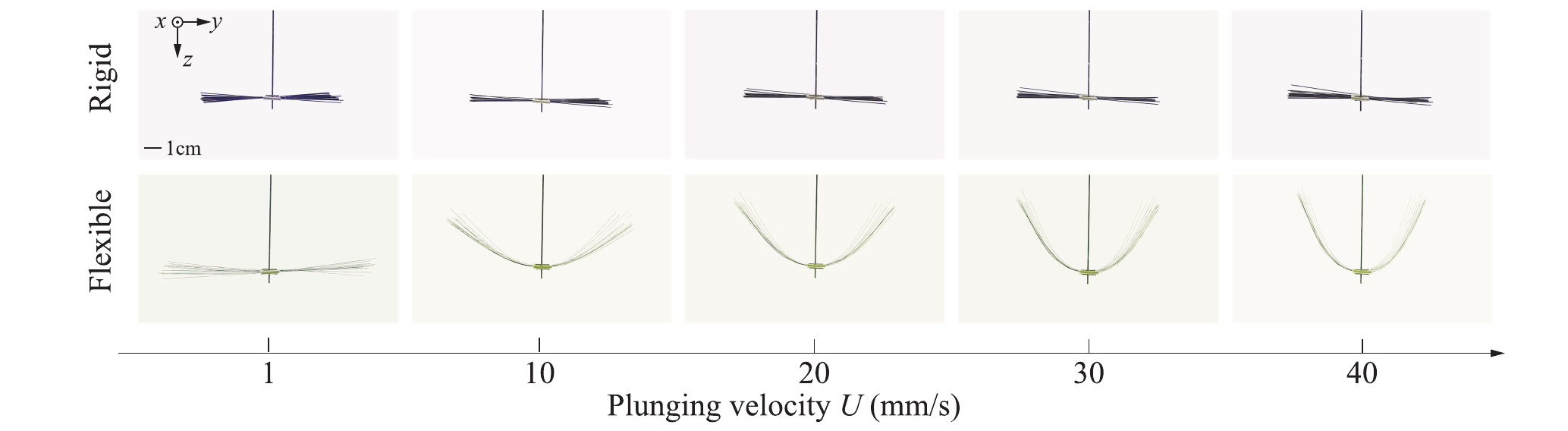}
        \caption{\textbf{Photographs (lateral view) of the experimental bristled strips driven through the viscous fluid.} Top row: Sample from set \textbf{A}, over a range of plunging velocities; parameters: $l{=}\SI{3}{\centi\meter}, D{=}\SI{0.58}{\milli\meter}$.
        Bottom row: Sample from set \textbf{C}; parameters: $l{=}\SI{6}{\centi\meter}, D{=}\SI{0.12}{\milli\meter}$.
        The scale bar (\SI{1}{\centi\meter}) and axis are the same for all panels. See Table~\ref{table:samples} for details on the sample parameters.
    }
        \label{fig:deformable_bristles}
    \end{figure}

The reconfiguration of flexible structures in flow leads to drag reduction and is often advantageous in many organisms~\cite{vogel1989, vogel1996}.
The examples in Fig.~\ref{fig:deformable_bristles} exhibit two types of qualitatively distinct behavior observed when plunging the bristled strips through the fluid bath: the bristles can either remain rigid or reconfigure.
The bristles in the sample sets \textbf{A} and \textbf{B} remain rigid during the plunge, even at the highest plunging velocity of 40 \unit[per-mode = symbol]{\milli\meter\per\second} (Fig.~\ref{fig:deformable_bristles}, top panel).
By contrast, the bristles in sample set \textbf{C} undergo reconfiguration, with significant deformation at velocities as low as 10 \unit[per-mode = symbol]{\milli\meter\per\second} (Fig.~\ref{fig:deformable_bristles} bottom row).
In this case, the drag coefficient depends on the deformed configuration, which is influenced by the combined effects of
the flow velocity, sample geometry, and the bending stiffness of the bristles: $F_\mathrm{drag} {=} \hat{F}_\mathrm{drag} (l, D, n, U, EI)$, $C_d {=} \hat{C}_d (l, D, n, U, EI)$~\cite{gosselin2010, alben2002, marchetti2018}, where $E$ is Young's modulus, $I {=} {\pi r^4}/{4}$ is the second moment of area, and $\hat{F}, \hat{C}_d$ denote functions.

\begin{figure}[ht!]
        \centering 
        \includegraphics[width=\textwidth]{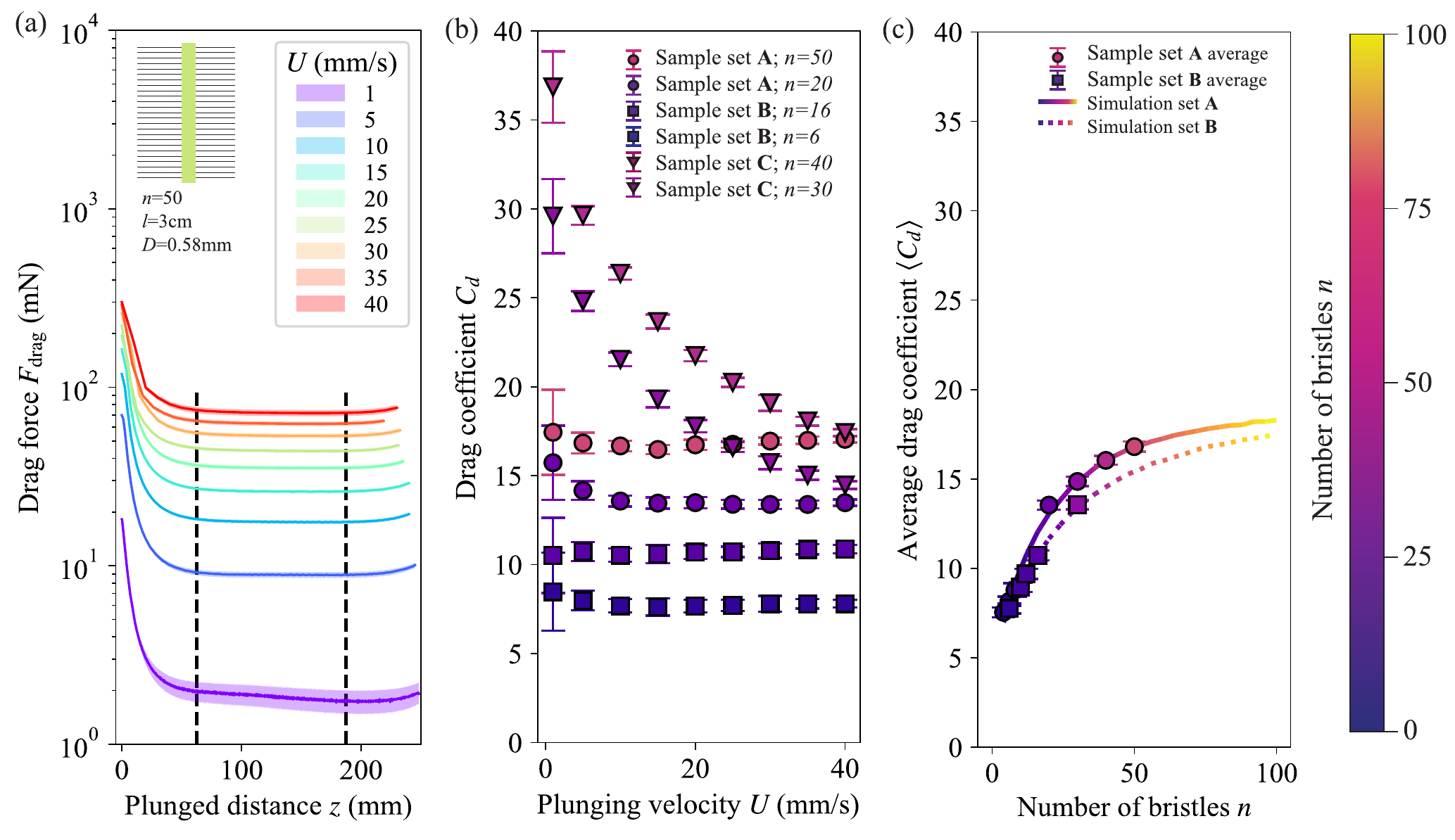}
        \caption{\textbf{Drag force and coefficient data: experiments vs.\ simulations.} (a) Representative examples of calibrated force signals for a bristled strip from set \textbf{A} with $n{=}50$. Inset: Schematic top view of the sample. 
        The solid lines are the averages, and the shaded regions represent ${\pm} 1$ standard deviation, with colors corresponding to plunging velocity $U$ (see legend). Average drag forces are computed using data within the two vertical dashed lines.
        (b) Drag coefficient $C_d$ of six bristled strips as a function of $U$. Symbols indicate the sample types (see legend), and colors represent the number of bristles (see the adjacent color map).
        (c) Experimental averages $\langle C_d \rangle$ of the rigid bristles strips (sample sets \textbf{A} and \textbf{B}), excluding data at 1 mm/s.
        Corresponding simulation results are plotted as solid and dashed lines, color-coded by the number of bristles, as in (b).
    }
        \label{fig:isolated_drag}
    \end{figure}

In Fig.~\ref{fig:isolated_drag}, we plot the drag coefficients $C_d$ of each bristled strip at different values of the plunging velocity $U$.
The raw drag-force signals as a function of the plunged distance $z$, at several values of $U$, are shown in Fig.~\ref{fig:isolated_drag}a for a representative sample.
The initial transient in the $F_\text{drag}(z)$ signals is due to the acceleration ramp-up of the drive and the interactions with the free surface of the fluid. At the end of each test, interactions between the sample and the bottom of the tank are also present. Thus, we discard the first and last quarters of each time series.
In between,  the approximately constant $F_\text{drag}(z)$ signal (within the vertical dashed lines in Fig.~\ref{fig:isolated_drag}a) is averaged for each bristled strip at a given $U$ to obtain $C_d$.
In Fig.~\ref{fig:isolated_drag}b, we plot $C_d$ versus $U$ for six selected samples, with different values of $l$, $D$, and $n$, observing a clear distinction between the rigid and reconfiguring samples. In the rigid cases (sets \textbf{A} and \textbf{B}, circle and square symbols, respectively), $C_d$ remains constant, independently of $U$.
By contrast, for the two samples from set \textbf{C} (triangles), $C_d$ decreases with $U$ due to the increasing bristle deformation (see Fig.~\ref{fig:deformable_bristles} bottom row).

For the rigid samples, we averaged the $C_d$ data from Fig.~\ref{fig:isolated_drag}b across a range of $U$ to obtain a mean value $\langle C_d\rangle$ for each bristled strip, which we plotted in Fig.~\ref{fig:isolated_drag}c as a function of the number of bristles $n$. 
Measurements taken at 1 mm/s are excluded due to the low signal-to-noise ratio.
The drag coefficients for all the samples in Fig.~\ref{fig:isolated_drag}c (circle and square symbols) increase with $n$, but with a lower rate of increase at larger values of $n$.
The results from the corresponding COMSOL simulations  (solid and dashed lines) are in excellent agreement with the experimental data (symbols).
Extending the number of bristles in simulations up to $n=100$ indicates that $C_d$ tends to approach a plateau.
The reconfiguration observed in set \textbf{C} samples and their non-constant $C_d(U)$ behavior raise the question of whether biological bristles undergo deformation during flight, which we tackle next.

\subsection{Scaling of drag force versus elasto-viscous number}
\label{sec:result:scaling}

The bristle geometry (i.e., aspect ratio and cross-sectional shape and area) affects the drag force experienced by each sample.
In the experiments, we observed significant reconfiguration in the samples with bristle length $l{=}\SI{6}{\centi\meter}$ and diameter $D{=}\SI{0.12}{\milli\meter}$, while those with shorter bristles ($l{=}\SI{3}{\centi\meter}$) and larger diameters ($D{=}\{0.17, 0.58\}$mm) remain undeformed (Fig.~\ref{fig:deformable_bristles}). 
The reconfiguration of a flexible structure can lead to a significant reduction in drag~\cite{vogel1989, vogel1996, alben2002, marchetti2018}, as evidenced by the $C_d$ values of bristled strips in sample set \textbf{C} (Fig.~\ref{fig:isolated_drag}b).
Drag reduction, even if desirable in many organisms, is disadvantageous for miniature insects that leverage drag  for flight~\cite{jones2015, farisenkov2022}.
It has been reported that bristled wings remain rigid or deform infinitesimally during flight~\cite{zhao2019} due to the high stiffness of the bristles~\cite{jiang2022}.
However, as evident by the results in Section~\ref{sec:result:rigid_v_flex}, we must also consider the fluid loading on the bristles together with their bending stiffness.

We proceed with a dimensional analysis to characterize the effect of reconfiguration on the drag force and enable quantitative comparisons between experimental and biological data.
We consider each bristle as a slender elastic fiber with a constant circular cross-section and bending stiffness $EI$.
The fiber deflection depends on the balance between the elastic and viscous forces.
In turn, the viscous drag depends on the shape of the fiber in flow~\cite{alben2002,marchetti2018}.
For simplicity, we consider a single fiber clamped on one end and moving with velocity $U$ in an infinite fluid domain (with )mass density $\rho$ and viscosity $\mu$).
Balancing the characteristic elastic and viscous forces, $EI/l^2{\sim}\mu U l$, one can construct the following elasto-viscous number (effectively, a dimensionless velocity)~\cite{marchetti2018}:
\begin{equation}
    \mathcal{V} = \frac{\mu U l^3}{EI}.
\end{equation}
Assuming that the characteristic curvature of the fiber is $\kappa {\sim} l^{-1}$ and balancing the moments due to the elastic and drag forces, $EI/l{\sim} F_\mathrm{drag}l$, we construct the dimensionless drag,
\begin{equation}
    \mathcal{D} = \frac{F_\mathrm{drag}l^2}{EI}.
\end{equation}
In low Re flow, when both the flow velocity and the fiber deflection are small, the scaling of drag with velocity is similar to that of a rigid fiber: $\mathcal{D} {\sim} \mathcal{V}$~\cite{marchetti2018}.
In the reconfiguration regime where the shape of the fiber is a function of the flow velocity, the scaling is expected to be $\mathcal{D} {\sim} \mathcal{V}^{1/2}$~\cite{marchetti2018}.
When the fiber reaches a limiting shape at even higher flow velocity, the scaling resumes to be $\mathcal{D} {\sim} \mathcal{V}$~\cite{marchetti2018}.
For pressure-dominated high Re flows, a similar decrease in the scaling exponent between  $\mathcal{D}$ and $\mathcal{V}$ is observed in the reconfiguration regime~\cite{alben2002, gosselin2010}.

\begin{figure}[ht]
        \centering 
        \includegraphics[width=\textwidth, keepaspectratio]{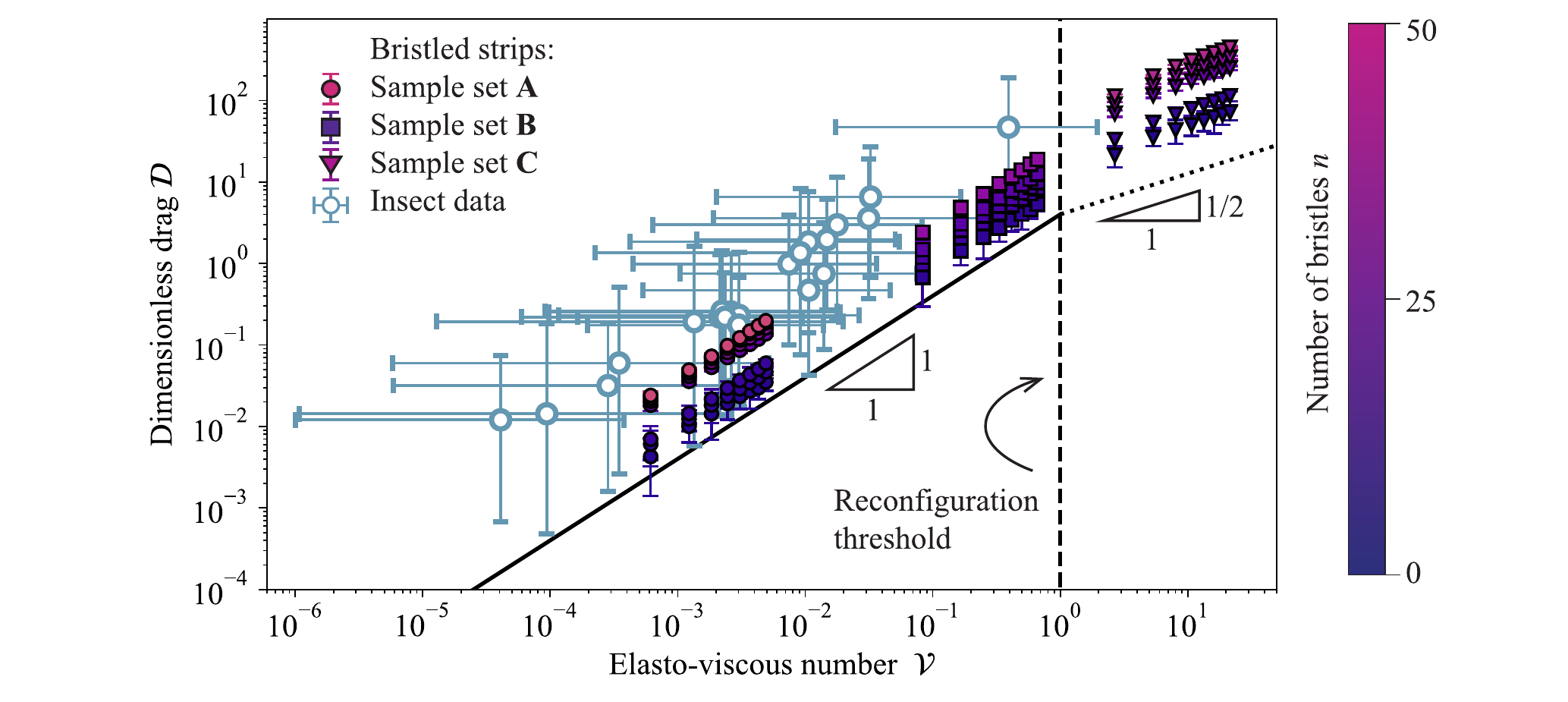}
        \caption{\textbf{Phase diagram for $\mathcal{D}$ and $\mathcal{V}$.} The dimensionless drag, $\mathcal{D}$ is plotted as a function of the elasto-viscous number, $\mathcal{V}$, for the experimental data (solid symbols: circles--sample set \textbf{A}, squares--sample set \textbf{B}, triangles--sample set \textbf{C}) and the estimated insect data (hollow circles). The experimental data points are color-coded by their value of $n$; see the adjacent color bar.
        The vertical dashed line marks $\mathcal{V}=1$. 
    }
        \label{fig:collapse_data}
    \end{figure}

We revisit the experimental data (Fig.~\ref{fig:isolated_drag}), together with the available insect morphology and flight data (Fig.~\ref{fig:definition}b), and apply the preceding dimensional analysis. The results are shown in Fig.~\ref{fig:collapse_data}.
Details on estimating biological quantities, such as the drag force and wing tip velocity, are provided below.
In computing $\mathcal{D}$ for the bristled strips, we neglect the drag force due to the bare (without bristles) central core, which was measured separately using the experimental protocol described in Section~\ref{sec:mats_methods:exp}.
This analysis does not include measurements at $U{=}1$mm/s.
Focusing first on the experiments, the data in Fig.~\ref{fig:collapse_data} exhibits two separate scaling regimes, in agreement with our previous observations in Section~\ref{sec:result:rigid_v_flex}.
The samples with short and thicker bristles ($l{=}\SI{3}{\centi\meter}$, in the sample sets \textbf{A} and \textbf{B}) remain undeformed even as $\Rey_L$ reaches the upper limit achievable with our experimental apparatus $(\Rey_L{=}4)$.
Hence, the scaling $\mathcal{D} {\sim} \mathcal{V}$ is observed for these samples (solid line in Fig.~\ref{fig:collapse_data}).
Large deflections in the bristles are observed in the sample set \textbf{C} ($l{=}\SI{6}{\centi\meter}$), even for $\Rey_L{\approx}1$. 
For these samples, $\mathcal{D}$ scales sub-linearly with $\mathcal{V}$.
Overall, we find that the results are consistent with the $\mathcal{D} {\sim} \mathcal{V}^{1/2}$ prediction~\cite{marchetti2018} (dotted line in Fig.~\ref{fig:collapse_data}), even if we do not attempt to fit the data to extract a precise exponent because the walls and inter-bristle interactions can both affect the expected scaling.

Using biological data available in the literature, we estimate $\mathcal{D}$ and $\mathcal{V}$ for 21 species of miniature insects~\cite{ huber2013, farisenkov2022, lin2007, huber2006, huber2007, huber2008, jiang2022, cheng2018}. We assumed they could at least hover and support their own body weight, which we used as an estimate for the minimal drag force their wings generate.
We could only find two reports of actual miniature insect weight~\cite{farisenkov2020, cheng2018}. Thus, we used these as minimal and maximum estimates for all insect species we shall consider.
For the wing tip velocity, we either directly used the values reported in the literature~\cite{cheng2018, jiang2022} or estimated it using supplemental videos of existing studies~\cite{farisenkov2022}.
Details of the estimation and insect data can be found in the SI.

The overlap between the $\mathcal{V}$ range in the insect data and that of the experimental samples in Fig.~\ref{fig:collapse_data} confirms that our experiments covered a biologically relevant parameter space.
The large errorbars of the biological data are due to the combined effect of data uncertainty, e.g., insect weights and bristle Young's modulus~\cite{farisenkov2020, jiang2022, cheng2018} and the geometry of individual wings, e.g., bristle length varies along the wing pad.
For the insect data, we could not estimate or subtract the drag force due to the wing pad alone.
It is important to note that all insect data lie below the reconfiguration threshold $\mathcal{V} {\sim} \mathcal{O}(1)$ (the vertical dashed line in Fig.~\ref{fig:collapse_data} marks $\mathcal{V} {=}1$), and the average values are consistent with a linear scaling.
Taken together, these results suggest that given the combined effect of viscous drag and elasticity in flapping flight, the biological bristled wings can avoid reconfiguration in their bristles, thus preventing drag reduction.
These findings also confirm that FSI simulations using a rigid-body assumption are appropriate for the subsequent numerical analysis.

\subsection{Optimal bristle configuration for maximum weighted drag}
\label{sec:result:sim_data}

For the rigid samples (from sets \textbf{A} and \textbf{B}), we find that the $\langle C_d \rangle$ data plotted in Fig.~\ref{fig:isolated_drag}c increase monotonically and approach a plateau.
This observation raises the hypothesis that there may exist an optimal number of bristles that maximizes $C_d$ when taking into account the weight of the bristles.
To test this hypothesis, we conduct thorough sweeps in the geometric-parameters space using the COMSOL simulations introduced in Section~\ref{sec:mats_methods:num}.

We now aim to find a parameter combining the drag generation and the weight of the bristled strips (or wings).
In the biological system, we assume that the wing pad and the bristles are made of the same material~\cite{jiang2022}.
Therefore, as a proxy for added weight, we consider the normalized total volume, $ \overline{V}$, defined as the ratio between the total volume of the bristled strip and the central core:
\begin{equation}
     \overline{V}  = \frac{n \pi D^2 l / 4 + BLt} {BLt}
     = 1 +  \frac{n \pi D^2 l} {4BLt}.
    \label{eq:vadd}
\end{equation}
Next, we define the \textit{weighted} Stokes drag coefficient as 
\begin{equation}
    \Cdbar = \frac{\Cd}{ \overline{V}} ,
    \label{eq:cd_bar}
\end{equation}
which captures both the drag force generated by the bristled strip and the volume of the added material.
We use $\Cdbar$ to assess the performance of bristled strip designs in the numerical studies described next to gain insight into the morphology of the bristled wings. 

In the simulations, we systematically varied the length  $l{\in}\{0.25, 1, 2, 3, 4, 5\}\,$cm, diameter $D\in\{0.4, 0.5, 0.6, 0.8, 1.0, 1.2\}\,$mm, and number $n{\in}[4,124]$ of the bristles, while ensuring $\mathcal{V}{\ll}1$ throughout.
The simulations were performed at $\Rey_L{=}1$, a choice that is consistent with the lower end of the reported ranges in the literature for biological bristled wings $\mathcal{O}(1){-}\mathcal{O}(10)$~\cite{sunada2002, farisenkov2020, wu_aerodynamics_2021}.

\begin{figure}[htbp]
        \centering 
        \includegraphics[width=\textwidth, keepaspectratio]{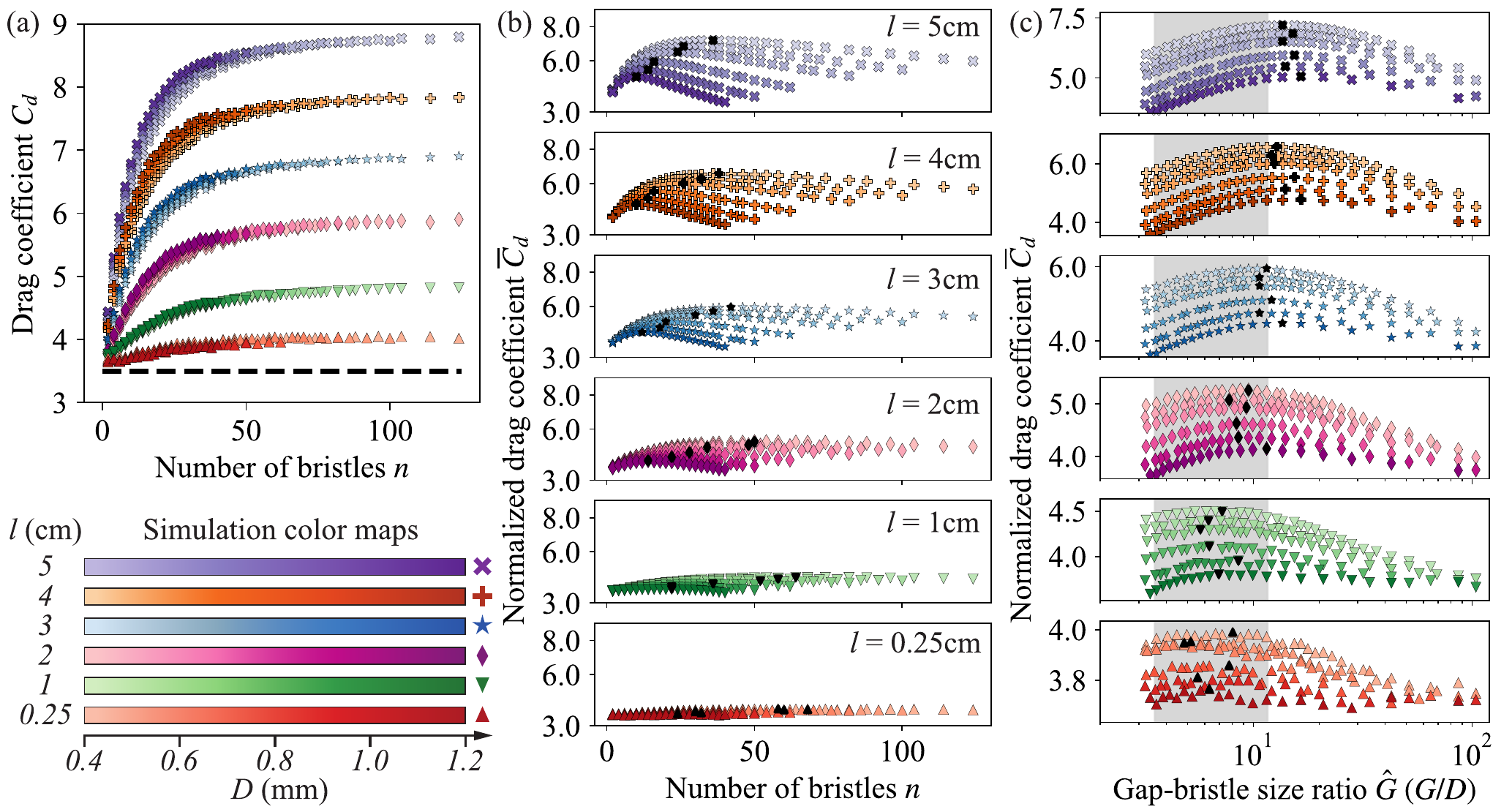}
        \caption{\textbf{Number of bristles to maximize weighted Stokes drag coefficients.}
        (a) Plot of $\Cd$ as a function of the number of bristles, simulated at $\Rey_L{=}1$.
        The horizontal dashed line represents the $\Cd$ value of the central core alone.
        As indicated by the color map (bottom left), each color series represents a set of simulations with a given bristle length, and the intensity of the color indicates bristle diameter. The diameter $D{\in}[0.4,1.2]$ mm  for all series. 
        (b) $\Cdbar$ values are plotted against the number of bristles $n$. The black symbol marks the location of $n^*$, which maximizes $\Cdbar$ in each data series.
        (c) $\Cdbar$ plotted against $\hat G {=}G/D$. Colors and symbols are the same as in (a,b).
        The grey-shaded regions correspond to the $\hat G$ range found in the literature for insect data (see SI Table S1).
    }
        \label{fig:sim_series_Re1}
    \end{figure}

The simulation results are shown in Fig.~\ref{fig:sim_series_Re1}. As indicated in the legend, different values of $l$ and $D$ and represented by the corresponding colors and their shades, respectively.
In Fig.~\ref{fig:sim_series_Re1}a, we plot the computed drag coefficient, $C_d$, versus the number of bristles, $n$.
The horizontal dashed line represents $C_d$ for the central core alone (without bristles).
The value of $C_d$ first increases sharply with $n$ and then reaches a plateau.
The behavior is qualitatively similar for strips with longer bristles, with higher plateaus for larger $l$ values.
For a fixed $l$, the diameter of the bristles influences the rate at which $C_d$ reaches the plateau, but this effect is relatively minor.

In Fig.~\ref{fig:sim_series_Re1}b, we plot $\Cdbar$ against $n$, finding that each data set with fixed $\{l,\, D\}$ and varying $n$ exhibits non-monotonic behavior, with a maximum in $\Cdbar$ at $n{=}n^*$, the optimal bristle number (marked by black symbols).
Consistently with what we have hypothesized, we observe that for fixed values of $l$ and $D$, as the number of bristles increases, $\Cdbar$ initially increases because the benefit of the added drag due to additional bristles exceeds the penalty due to the additional weight. 
As the bristle number further increases, the additional weight penalty outcompetes the benefit of the additional drag, causing $\Cdbar$ to decrease.
Maximizing $\Cdbar$ represents the best balance between gaining additional drag and taking on more weight by adding more bristles.
Additionally, the effect of the bristle volume is highlighted in Fig.~\ref{fig:sim_series_Re1}b: for fixed $n$ and $l$, the bristles strips with lower $D$ generate a higher value of $\Cdbar$ because the bristles make smaller contributions to the total volume $\overline V$, Eq.~\eqref{eq:vadd}, but relatively similar contributions to generated drag (Fig.~\ref{fig:sim_series_Re1}a).

The inter-bristles gap size is an important parameter, determining the type of flow between the bristles and the leakiness of the bristled wing~\cite{cheer1987, kasoju_leaky_2018, lee_optimal_2020, kasoju_interspecific_2021, lee_unsteady_2018, lee_gusty_2021}.
We define the dimensionless gap size $\hat G{=}G/D$, where $G$ is the distance from the edge of one bristle to that of its neighbor.
In Fig.~\ref{fig:sim_series_Re1}c, we replot the $\Cdbar$ data from Fig.~\ref{fig:sim_series_Re1}b, now as a function of $\hat G$.
We observe that the $\Cdbar$ maxima in a simulation set tend to cluster, and $\hat G {\in} [4.8, 16.5]$ over all the datasets. 
These clusters of maxima are within or close to the range of data ($\hat G_\text{bio} {\in} [3.6, 11.6]$, indicated by the grey-shaded region in Fig.~\ref{fig:sim_series_Re1}c) for biological bristled wings reported in the literature (see SI Table S1).
While the maxima for shorter bristles lengths fall within the biological $\hat G$ range, the maxima for longer bristles ($l {=} 4{-}\SI{5}{\milli\meter})$ are larger than the upper end of the biological range. 
The overestimation of $\hat G$ at large $l$ values could be a result of the simulated kinematics and the usage of total wing volume, Eq.~\eqref{eq:vadd}, in the optimization objective function, Eq.~\eqref{eq:cd_bar}.
Moreover, the mismatch between simulation and biological data could also suggest that factors besides drag generation contribute to the observed wing morphology.
For example, a possible safety margin in bristle diameter to avoid bristle reconfiguration would lead to larger $D$ and smaller $\hat G$ in insect wings.
More broadly, aspects not considered in this study, such as overall wing geometry, more complex flight kinematics, and power consumption of flapping, most likely have also influenced the evolution of wing morphology.

\begin{figure}[ht]
        \centering 
        \includegraphics[width=\textwidth, keepaspectratio]{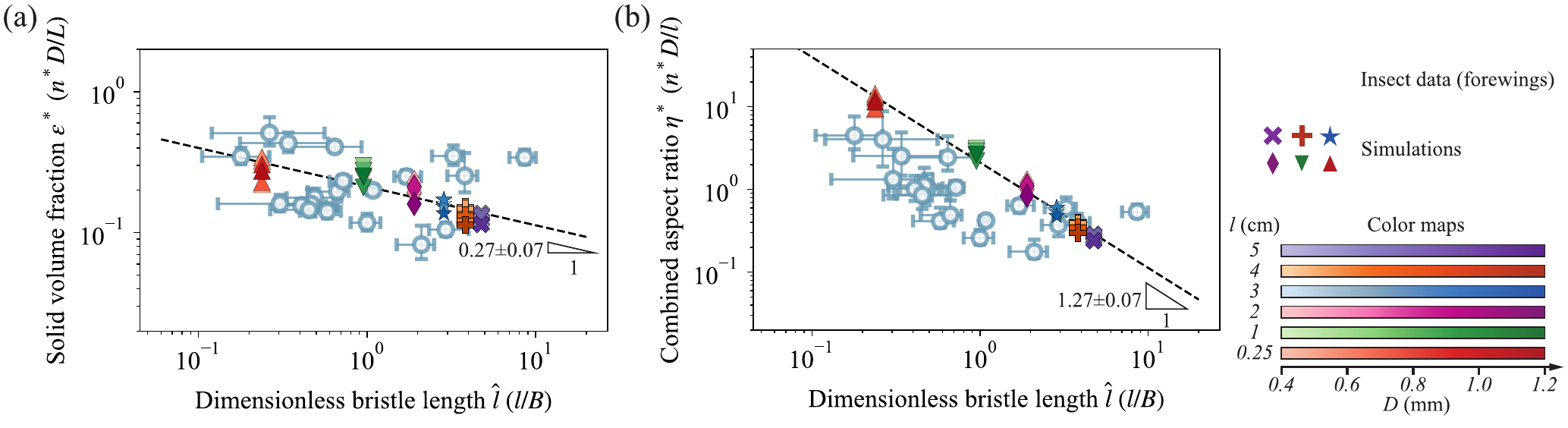}
        \caption{
        \textbf{Optimal solid fraction $\epsilon^*$ and combined aspect ratio $\eta^*$ versus bristle length.}
        (a) The solid volume fraction $\epsilon^* {=} n^*D/L$ is plotted against $\hat l {=} l/B$, the ratio between the length of the bristles and the width of the strip. For insect data, average wing pad length $L$ and width $B$, and average bristle length $l$ are used in the calculation of $\epsilon^*$.
        Both simulation data (solid symbols) and insect data (open circles with error bars) are shown, with a colormap identical to that of Fig.~\ref{fig:sim_series_Re1}.
        The dashed line corresponds to the power-law $\epsilon^*{\sim}\hat{l}^\gamma$, with an exponent
        $\gamma {=} -0.27{\pm}0.07$ obtained from a least-squares fit of the simulation data. The $\pm$ uncertainty corresponds to the 95\% confidence interval of the fit.
        (b) The combined aspect ratio $\eta^* {=} n^*D/l$ is plotted against $\hat l$.
        The dashed line corresponds to the power-law $\eta^*{\sim}\hat{l}^{\gamma-1}$.
        Symbols and colormaps are identical to (a).
    }
        \label{fig:optimal_Re1}
    \end{figure}

In Fig.~\ref{fig:optimal_Re1}, we compare optimal configurations found in simulations with the existing biological data.
First, we define the bristle solid volume fraction $\epsilon$ by taking the ratio between the total bristle volume and the enclosing volume (considering only bristles on one side of the central core due to symmetry):
\begin{equation}
    \frac{\text{Total bristle volume}}{\text{Enclosing volume}} =  \frac{\frac{n}{2} \pi (D/2)^2 l}{DLl} \sim \frac{nD}{L} \equiv \epsilon.
\end{equation}
Solid volume fraction, or equivalently, porosity, is a  parameter that is commonly used when studying porous systems, such as an array of cylinders \cite{tamada_steady_1957, kirch1962, drummond_laminar_1984} and porous discs and strips \cite{strong_2019, pezzulla_2020}.
In Fig.~\ref{fig:optimal_Re1}a, we plot $\epsilon^*$ (closed symbols), which is evaluated with the optimal bristle number $n^*$, as a function of $\hat l {=} l/B$, where $B$ is the width of the central core or the wing pad.
In addition, we overlay the corresponding biological data (open symbols) from Fig.~\ref{fig:definition}b (Table S1).
We observe that $\epsilon^*$ depends on the dimensionless bristle length $\hat l$, and their scaling relationship is consistent the power-law $\epsilon^*{\sim}\hat{l}^\gamma$, with an exponent $\gamma{=-}0.27{\pm}0.07$, where the uncertainty corresponds to the 95\% confidence interval of the least-squares fit.
The biological data, on the other hand, exhibit natural variations and large uncertainties.
However, despite the scatter, the biological data follows the trend of the simulated data.

Next, we define an alternative parameter $\eta^* {=} n^*D/l$, which can be interpreted as the inverse aspect ratio of the combined projected area of the bristles (combined aspect ratio for short). 
In low Re number flow past an array of \textit{finite} cylinders, the fluid tends to flow around the array, leading to a relatively small drag on the interior cylinders \cite{tamada_steady_1957}.
Therefore, the drag on the array can be mostly attributed to the shear layers concentrated at the edge of the bristle strip and at the bristle tips.
However, shear layers can also develop in the region in the gap near the tip, which could, in turn, depend on the bristle length.
Due to this physical intuition, the combined aspect ratio $\eta^*$, which considers the geometry of the combined projected area, may be relevant in interpreting the optimal bristle configurations.
In Fig.~\ref{fig:optimal_Re1}b, we replot the data from Fig.~\ref{fig:optimal_Re1}a but with $\eta^*$ on the $y$-axis.
The scaling relation now becomes $\eta^* {\sim} \hat l^{\gamma-1}$ since we have gained an additional $\hat l^{-1}$ in the definition of $\eta^*$. 
Interestingly, except for two outliers, all biological data lie below the upper bound provided by the simulations data.
Taken together, these results suggest that the total perimeter of the bristle tips $n^*D$ (at optimal simulated configuration or in biological wings) may be a scale-invariant function of the bristle length $l$.
However, more precise and comprehensive biological data sets would be required to ascertain the suggested morphological trend conclusively.

\section{Discussion}
\label{sec:dis}

The overall agreement between the predictions from our analyses and the actual biological data (Figs.~\ref{fig:collapse_data} and \ref{fig:optimal_Re1}) speaks to the value of our combined  experiments and computations using reduced models.
Despite this success, our study also has several limitations.
Due to a lack of data, we had to use values collected or estimated on a few species of insects (e.g., weight and wing tip velocity~\cite{cheng2018, jiang2022, farisenkov2022}) as substitutions for other species, resulting in large uncertainties in the data presented in Fig.~\ref{fig:definition}b, Fig.~\ref{fig:collapse_data}, and Fig.~\ref{fig:optimal_Re1}.

Experimentally, with its rectangular design for the bristled strips, the reduced physical model ignored the diverse shapes found in biological wings (Fig.~\ref{fig:definition}a), thus leaving a large design space unexplored. 
The experiments also neglected the complex flight kinematics used by bristled-winged insects~\cite{weis-fogh1975, jones2015, cheng2018, shen2022}.
However, as we have mentioned in Section~\ref{sec:prob_def}, our results are relevant to drag-based flight mechanisms using bristled wings, given that drag at high angles of attack is the main source of lift and thrust~\cite{farisenkov2020, engels_flight_2021, farisenkov2022, shen2022}.


\section{Conclusion}
\label{sec:conclusion}

We combined experiments and simulations of a reduced model of bristled wings to investigate the role of bristle geometry on drag generation and to attempt a rationalization of trends observed in the biological data (Fig.~\ref{fig:definition}).
After validating the simulations with experimental data, we leveraged the numerics to systematically explore the parameter space, including bristle length, diameter, and number.
Our main results, presented in Figs.~\ref{fig:collapse_data} and~\ref{fig:optimal_Re1}, offer several insights that supplement prior research on bristled wing morphologies~\cite{lee_optimal_2020, kasoju_interspecific_2021, engels_flight_2021}.
First, avoiding reconfiguration of the bristles during flight appears to be a general feature in the biological wings we examined (Fig.~\ref{fig:collapse_data}).
As such, the trade-off between increasing drag generation and increasing weight by adding more bristles to a wing may be an important factor in the evolution of the morphology of the bristled wings of miniature insects (Fig.~\ref{fig:optimal_Re1}).

Moreover, the normalized drag coefficient, $\Cdbar$, introduced in Eq.~\ref{eq:cd_bar}, encapsulates two essential factors influencing wing function and serves as an effective performance indicator for the bristled strips (wings).
We found a scaling relationship between the solid volume fraction $\epsilon^*{=} n^*D/L$, where $n^*$ maximizes $\Cdbar$ given fixed ${l, D}$, and dimensionless bristle length $\hat l{=}l/B$.
The simulation data is consistent with power-law $\epsilon^*{\sim}\hat{l}^\gamma$ with $\gamma{=-}0.27{\pm}0.07$ (Fig.~\ref{fig:optimal_Re1}a).
Similarly, we found a scaling relationship between the combined aspect ratio $\eta^* {=} n^*D/l$ and $\hat l$ (Fig.~\ref{fig:optimal_Re1}b).
While we did provide a rationale for the $\mathcal{D}{\sim}\mathcal{V}$ scaling in the biological data (Fig.~\ref{fig:collapse_data}), we have been unable to construct an argument to support this potential scaling between $\epsilon^*$ and $\hat l$ (or between $\eta^* $ and $\hat l$).
We note that even though the biological data has a considerable scatter due to a combination of natural variations and estimation uncertainties, it follows the trend in the simulation data $\epsilon^*(\hat l)$ (Fig.~\ref{fig:optimal_Re1}a).
Additionally, the $\eta^*(\hat l)$ simulation data offers an upper bound for the insect data (Fig.~\ref{fig:optimal_Re1}b). 
We hope our suggestive results will motivate more formal theoretical analyses in future research.
Regardless, the proposed quantitative characterization, such as the scaling between $\mathcal{D}$ and $\mathcal{V}$, and $\Cdbar$, can guide the design of bio-mimetic applications.

In the dimensional analysis in our study (Section~\ref{sec:result:scaling}), we have neglected the inter-bristle hydrodynamic coupling and other potential effects on the collective bending stiffness of the bristles, which are also interesting aspects to explore in future work.
In general, the mathematical modeling of bristled wings in flow remains understudied due to the inherent theoretical challenges.
A description of bristled regions in flow as a homogenized porous material has been recently proposed~\cite{santhanakrishnan_clap_2014}; however, additional work is needed to integrate and examine the effects of elasticity and geometric nonlinearities.
Other studies on poroelastic structures have revealed surprising and insightful results in both experimental and biological contexts~\cite{strong_2019, pezzulla_2020, cummins2017, cummins2018, ledda2019}. 
Similar approaches could be adopted to investigate the FSI of bristled wings in future research.

\section{Acknowledgement}
We express our gratitude to Pierre Leroy-Calatayud for his assistance with preliminary experiments during the early phases of this research.

\newpage
\printbibliography

\end{document}


\maketitle

\section{Morphological data}
\subsection{Measurements of bristled forewings of 21 species of insects.}

In Section 4.2 of the main text, we present scaling between dimensionless drag $\mathcal{D} = F_\mathrm{drag} l^2/EI$ and the elasto-viscous number $\mathcal{V} = \mu U l^3 / EI$ and compare experimental and biological data.
To calculate the dimensionless drag $\mathcal{D}$, we used the reported weight of miniature insects in refs~\cite{farisenkov2022,cheng2018} as values of $F_\mathrm{drag}$.
The values of Young's modulus $E$ are taken from ref.~\cite{jiang2022}. The second moment of area can be computed from values of bristle diameter $D$, which we assume to be constant throughout the entire bristle. 
The velocity $U$ in the elasto-viscous number $\mathcal{V}$ is estimated using the wing tip velocity or wing beat frequencies reported by refs~\cite{farisenkov2022, cheng2018, jiang2022}.
The viscosity is that of air at $20^\circ$C.
Values of $D$ and $G$ are obtained from supplementary information Table S1 of ref~\cite{jones2016}, except for \textit{P.~placentis}, which is measured directly from the insect forewing image in ref.~\cite{farisenkov2022} using ImageJ~\cite{schneider2012}.
All other quantities, i.e., bristle length $l$, number $n$, wing pad length $L$, and width $B$, are all measured directly from images in the corresponding references using ImageJ~\cite{schneider2012}.
Other measurement details can be found in the main text. The data on the insect forewings are documented in Table~\ref{table:insect_forewings} and can be found in supplemental datasheet \texttt{S1{\textunderscore}insect{\textunderscore}data.xlsx}.

\begin{table}[!ht]
    \centering
    \resizebox{\textwidth}{!}{%
    \begin{tabular}{|l|l|l|l|l|l|l|l|l|l|l|l|l|l|l|l|l|l|}
   \hline
        \textbf{Species and source} & \textbf{Figure} & \textbf{Wing pad shape} & $\mathcal{D}_\mathrm{min}$ & $\mathcal{D}_\mathrm{avg}$ & $\mathcal{D}_\mathrm{max}$ & $\mathcal{V}_\mathrm{min}$ & $\mathcal{V}_\mathrm{avg}$ & $\mathcal{V}_\mathrm{max}$ & $l_\mathrm{min}$ (mm) & ${l}_\mathrm{avg}$ (mm) & $l_\mathrm{max}$  (mm) & $D$ ($\mu$m) & $n$ & $L$ (mm) & $\Delta L$ (mm) & $B$ (mm) & $G$ ($\mu$m) \\ \hline
        \textit{Allanagrus magniclava}~\cite{lin2007} & Fig 14 & Elongated & 1.59E-03 & 3.18E-02 & 1.88E-01  & 5.91E-06 & 2.84E-04 & 2.51E-03  & 0.05 & 0.09 & 0.13 & 2.44 & 90 & 0.54 & 0.041 & 0.135 & 9.00 \\ \hline
        \textit{Cleruchoides noackae}\cite{lin2007} & Fig 269 & Elongated & 8.79E-02 & 7.49E-01 & 3.19E+00  & 1.05E-03 & 1.40E-02 & 7.53E-02  & 0.16 & 0.188 & 0.23 & 1.6 & 75 & 0.48  & 0.0466 & 0.113 & 7.83 \\ \hline
        \textit{Dicopomorpha schleideni}~\cite{lin2007} & Fig 92 & Elongated & 3.68E-01 & 3.59E+00 & 1.90E+01  & 1.92E-03 & 3.14E-02 & 2.34E-01  & 0.07 & 0.088 & 0.12 & 0.74 & 44 & 0.31 & 0.0336 & 0.031 & 6.01 \\ \hline
        \textit{Eubroncus dubius}~\cite{lin2007} & Fig 123 & Oval & 6.76E-04 & 1.21E-02 & 7.42E-02  & 1.01E-06 & 4.08E-05 & 3.81E-04  & 0.02 & 0.034 & 0.05 & 1.91 & 80 & 0.44 & 0.0583 & 0.188 & 8.53 \\ \hline
        \textit{Eustochus besucheti}~\cite{huber2007} & Fig 27 & Oval & 1.86E-02 & 2.54E-01 & 1.31E+00  & 9.71E-05 & 2.63E-03 & 1.88E-02  & 0.07 & 0.104 & 0.14 & 1.56 & 58 & 0.58 & 0.0679 & 0.220 & 15.33 \\ \hline
        \textit{Eustochus nearticus}~\cite{huber2007} & Fig 28 & Oval & 1.74E-02 & 2.31E-01 & 1.36E+00  & 1.16E-04 & 3.03E-03 & 2.65E-02  & 0.09 & 0.132 & 0.19 & 1.8 & 77 & 0.78  & 0.104 & 0.277 & 14.6 \\ \hline
        \textit{Eustochus nipponicus}~\cite{huber2007} & Fig 34 & Oval & 1.33E-02 & 2.20E-01 & 1.27E+00  & 5.96E-05 & 2.14E-03 & 1.83E-02  & 0.06 & 0.098 & 0.14 & 1.57 & 70 & 0.69 & 0.151 & 0.225 & 14.04 \\ \hline
        \textit{Eustochus pengellyi}~\cite{huber2007} & Fig 30 & Oval & 2.05E-02 & 2.60E-01 & 1.44E+00  & 9.16E-05 & 2.23E-03 & 1.77E-02  & 0.06 & 0.086 & 0.12 & 1.41 & 62 & 0.56 & 0.0555 & 0.215 & 13.57 \\ \hline
        \textit{Eustochus yoshimotoi}~\cite{huber2007} & Fig 31 & Oval & 5.75E-03 & 1.93E-01 & 1.62E+00  & 1.28E-05 & 1.35E-03 & 1.99E-02  & 0.03 & 0.07 & 0.12 & 1.37 & 68 & 0.58 & 0.0775 & 0.259 & 12.34 \\ \hline
        \textit{Kikiki huna}~\cite{huber2013} & Fig 27 & Elongated & 9.95E-02 & 9.86E-01 & 3.93E+00  & 4.44E-04 & 7.45E-03 & 3.63E-02  & 0.06 & 0.076 & 0.09 & 0.95 & 32 & 0.12 & 0.0373 & 0.015 & 9.13 \\ \hline
        \textit{Mimalaptus victoria}~\cite{lin2007} & Fig 158 & Elongated & 4.23E-02 & 4.70E-01 & 1.74E+00  & 5.36E-04 & 1.07E-02 & 4.64E-02  & 0.17 & 0.228 & 0.26 & 1.98 & 69 & 0.39 & 0.0618 & 0.079 & 10.75 \\ \hline
        \textit{Mymaromella chaoi}~\cite{huber2008} & Fig 17 & Teardrop & 6.78E-01 & 6.54E+00 & 2.68E+01  & 2.02E-03 & 3.26E-02 & 1.65E-01  & 0.04 & 0.05 & 0.06 & 0.48 & 27 & 0.11 & 0.0149 & 0.047 & 2.65 \\ \hline
        \textit{Mymaromella cyclopterus}~\cite{huber2008} & Fig 19 & Teardrop & 1.42E-01 & 1.84E+00 & 7.62E+00  & 4.22E-04 & 1.06E-02 & 5.47E-02  & 0.04 & 0.06 & 0.07 & 0.71 & 34 & 0.17 & 0.0211 & 0.091 & 5.85 \\ \hline
        \textit{Mymaromella mira}~\cite{huber2008} & Fig 22 & Teardrop & 2.15E-01 & 2.98E+00 & 1.15E+01  & 6.39E-04 & 1.78E-02 & 8.29E-02  & 0.04 & 0.06 & 0.07 & 0.64 & 46 & 0.15 & 0.0166 & 0.089 & 4.81 \\ \hline
        \textit{Mymaromella pala}~\cite{huber2008} & Fig 20 & Teardrop & 2.69E-01 & 1.96E+00 & 6.18E+00  & 1.40E-03 & 1.48E-02 & 5.07E-02  & 0.07 & 0.076 & 0.08 & 0.8 & 40 & 0.16 & 0.0121 & 0.066 & 5.52 \\ \hline
        \textit{Prionaphes depressus}~\cite{lin2007} & Fig 223 & Oval & 2.46E-02 & 2.19E-01 & 7.68E-01  & 1.65E-04 & 2.35E-03 & 9.46E-03  & 0.09 & 0.108 & 0.12 & 1.65 & 69 & 0.49 & 0.0427 & 0.154 & 6.64 \\ \hline
        \textit{Schizophragma basalis}~\cite{lin2007} & Fig 242 & Oval & 7.59E-02 & 1.35E+00 & 8.33E+00  & 2.26E-04 & 9.16E-03 & 8.55E-02  & 0.04 & 0.068 & 0.1 & 0.83 & 70 & 0.4 & 0.0354 & 0.140 & 5.67 \\ \hline
        \textit{Stethynium breviovipositor}~\cite{huber2006} & Fig 9 & Oval & 2.6E-03 & 6.01E-02 & 5.08E-01  & 5.82E-06 & 3.47E-04 & 5.21E-03  & 0.03 & 0.058 & 0.1 & 1.67 & 88 & 0.34 & 0.0542 & 0.168 & 6.97 \\ \hline
        \textit{Stethynium ophelimi}~\cite{huber2006} & Fig 1 & Oval & 4.79E-04 & 1.43E-02 & 1.83E-01  & 1.07E-06 & 9.40E-05 & 2.63E-03  & 0.03 & 0.066 & 0.14 & 2.55 & 104 & 0.52 & 0.118 & 0.251 & 9.07 \\ \hline
        \textit{Tinkerbella nana}~\cite{huber2013} & Fig 5 & Elongated & 3.88E+00 & 4.70E+01 & 1.90E+02  & 1.74E-02 & 3.93E-01 & 1.95E+00  & 0.06 & 0.084 & 0.1 & 0.38 & 39 & 0.18 & 0.0480 & 0.028 & 4.4 \\ \hline
        \textit{Paratuposa placentis}~\cite{farisenkov2022} & Fig 1 & Elongated & 1.88E-02 & 1.76E-01 & 6.75E-01  & 1.96E-04 & 3.00E-03 & 1.39E-02  & 0.14 & 0.172 & 0.2 & 2.2 & 42 & 0.27 & 0.0324 & 0.020 & 18.1 \\ \hline
    \end{tabular}
    }
     \caption{Morphological measurements of bristled forewings of 21 species of insects.}
    \label{table:insect_forewings}
\end{table}

\subsection{Measurements of bristled hindwings of 15 species of insects.}
Similar to the measurements of forewings, all quantities, i.e., bristle length $l$, number $n$, wing pad length $L$, and width $B$, are measured directly from images of insect hindwings in the corresponding references using ImageJ~\cite{schneider2012}.
The figure number in the reference for each species is the same as documented in Table~\ref{table:insect_forewings}.
Other measurement details can be found in the main text. The data on the insect forewings are documented in Table~\ref{table:insect_forewings} and can also be found in supplemental datasheet \texttt{S1{\textunderscore}insect{\textunderscore}data.xlsx}.

\begin{table}[!ht]
    \centering
    \resizebox{\textwidth}{!}{%
    \begin{tabular}{|l|l|l|l|l|l|l|l|}
    \hline
        \textbf{Species and reference} & \textbf{Wing pad shape} & $n$ & $L$ (mm) & $B$ (mm) & ${l}_\mathrm{min}$ (mm) & ${l}_\mathrm{avg}$ (mm) & ${l}_\mathrm{max}$ (mm) \\ \hline
        \textit{Allanagrus magniclava}~\cite{lin2007} & Elongated & 76 & 0.48 & 0.047 & 0.04  & 0.06 & 0.10  \\ \hline
        \textit{Cleruchoides noackae}~\cite{lin2007} & Elongated & 49 & 0.43 & 0.044 & 0.06  & 0.13 & 0.17  \\ \hline
        \textit{Dicopomorpha schleideni}~\cite{lin2007} & Elongated & 30 & 0.49 & 0.023 & 0.11  & 0.17 & 0.21  \\ \hline
        \textit{Eubroncus dubius}~\cite{lin2007} & Oval & 79 & 0.42 & 0.078 & 0.03  & 0.05 & 0.08  \\ \hline
        \textit{Mimalaptus victoria} ~\cite{lin2007} & Elongated & 29 & 0.22 & 0.026 & 0.10  & 0.13 & 0.15  \\ \hline
        \textit{Prionaphes depressus} ~\cite{lin2007} & Elongated & 53 & 0.4 & 0.047 & 0.03  & 0.08 & 0.12  \\ \hline
        \textit{Schizophragma basalis} ~\cite{lin2007} & Elongated & 67 & 0.33 & 0.026 & 0.03  & 0.05 & 0.08  \\ \hline
        \textit{Kikiki huna}~\cite{huber2013} & Elongated & 16 & 0.12 & 0.005 & 0.04  & 0.06 & 0.07  \\ \hline
        \textit{Tinkerbella nana} ~\cite{huber2013} & Elongated & 26 & 0.13 & 0.010 & 0.03  & 0.07 & 0.11  \\ \hline
        \textit{Eustochus besucheti}~\cite{huber2007} & Elongated & 41 & 0.47 & 0.035 & 0.04  & 0.09 & 0.13  \\ \hline
        \textit{Eustochus nearticus}~\cite{huber2007} & Elongated & 44 & 0.62 & 0.033 & 0.06  & 0.10 & 0.14  \\ \hline
        \textit{Eustochus nipponicus}~\cite{huber2007} & Elongated & 49 & 0.55 & 0.037 & 0.03  & 0.08 & 0.14  \\ \hline
        \textit{Eustochus pengellyi}~\cite{huber2007} & Elongated & 43 & 0.46 & 0.028 & 0.04  & 0.07 & 0.11  \\ \hline
        \textit{Stethynium breviovipositor}~\cite{huber2006} & Elongated & 57 & 0.3 & 0.032 & 0.03  & 0.06 & 0.10  \\ \hline
        \textit{Stethynium ophelimi}~\cite{huber2006} & Elongated & 60 & 0.47 & 0.042 & 0.01  & 0.07 & 0.12  \\ \hline
    \end{tabular}
    }
    \caption{Morphological measurements of bristled hindwings of 15 species of insects.
    }
    \label{table:insect_hindwings}
\end{table}
\FloatBarrier

\section{Additional figure}

\begin{figure}[hbtp]
    \centering
    \includegraphics[width =\textwidth, keepaspectratio]{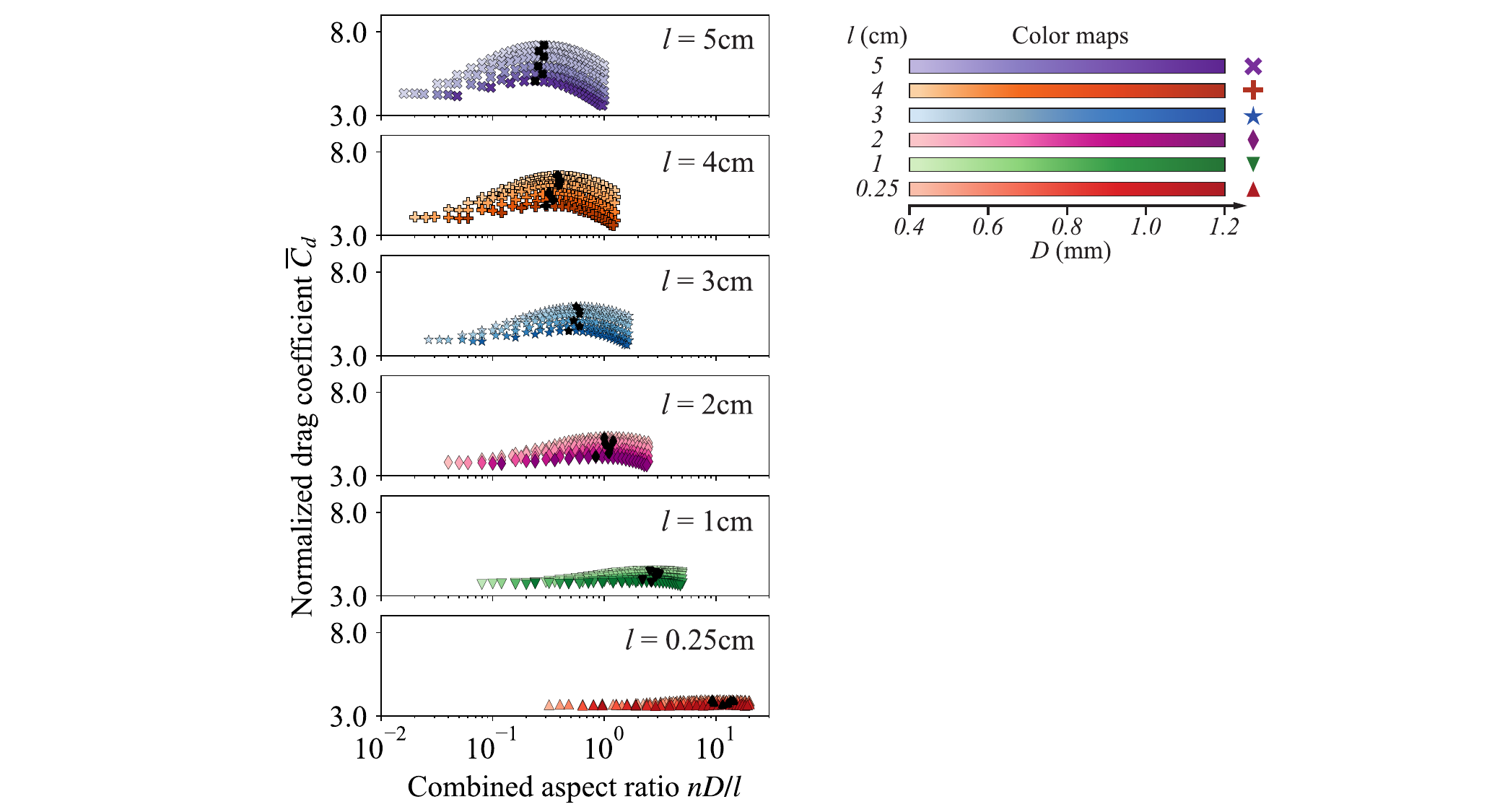}
    \caption{The normalized drag coefficient $\overline{C}_d$ is plotted against the inverse aspect ratio of the combined projected area of the bristles, $\eta=nD/l$.
    }
    \label{fig:sup1}
\end{figure}

\FloatBarrier

\section{Data for plotting figures in the manuscript}

We provide a supplemental dataset \texttt{S2{\textunderscore}data{\textunderscore}for{\textunderscore}all{\textunderscore}figures.xlsx} to reproduce the following figures in the manuscript:
\begin{enumerate}
    \item Fig. 1b
    \item Fig. 4a--c
    \item Fig. 5
    \item Fig. 6a--c
    \item Fig. 7a, b
    \item Supplementary Fig. S1
\end{enumerate}

\newpage
\printbibliography